\documentclass[10pt,twoside,spanish,english]{article}
\usepackage[T1]{fontenc}
\usepackage[latin9]{inputenc}
\usepackage[letterpaper]{geometry}
\usepackage{color}
\usepackage{babel}
\addto\shorthandsspanish{\spanishdeactivate{~<>.}}

\usepackage{float}
\usepackage{amsthm}
\usepackage{amsmath}
\usepackage{graphicx}
\usepackage{setspace}
\usepackage[unicode=true,pdfusetitle,
 bookmarks=true,bookmarksnumbered=true,bookmarksopen=true,bookmarksopenlevel=1,
 breaklinks=false,pdfborder={0 0 0},backref=false,colorlinks=false]
 {hyperref}
\usepackage{breakurl}

\makeatletter
\numberwithin{equation}{section}
\numberwithin{figure}{section}
  \theoremstyle{remark}
  \newtheorem*{acknowledgement*}{\protect\acknowledgementname}

\usepackage{multicol}

\date{} 

\makeatother

  \addto\captionsenglish{\renewcommand{\acknowledgementname}{Acknowledgement}}
  \addto\captionsspanish{\renewcommand{\acknowledgementname}{Agradecimientos}}
  \providecommand{\acknowledgementname}{Acknowledgement}

\begin{document}

\title{\noindent \textbf{Is the $\left(3+1\right)-d$ nature of the universe
a thermodynamic necessity?}}

\author{\textbf{Julian Gonzalez-Ayala and F Angulo-Brown.}}

\maketitle
\begin{spacing}{0.10000000000000001}
\noindent \begin{center}
\textsl{Departamento de F\'isica, Escuela Superior de F\'isica y
Matem\'aticas, Instituto Polit\'ecnico Nacional, }\\
\textsl{Edificio 9, CP 07738, M\'exico D. F. }\\
\textsl{E-mail: julian@esfm.ipn.mx, angulo@esfm.ipn.mx}\\

\par\end{center}
\end{spacing}
\begin{abstract}
\begin{singlespace}
\noindent It is well established that at early times, long before
the time of radiation-matter equality, the universe could have been
well described by a spatially flat, radiation only model. In this
letter we consider the whole primeval universe as a black body radiation
(BBR) system in an $n-$dimensional Euclidean space. We propose that
the $\left(3+1\right)-d$ nature of the universe could be the result
of a kind of thermodynamic selection principle stemming from the second
law of thermodynamics. In regard the three spatial dimensions we suggest
that they were chosen by means of the minimization of the Helmholtz
free energy per hypervolume unit, while the time dimension, as it
is well known was the result of the principle of increment of entropy
for closed systems; that is, the so-called arrow of time.\end{singlespace}

\end{abstract}
\begin{multicols}{2}

In 1989, Landsberg and de Vos \cite{key-11} proposed a spatial $n-$dimensional
generalization of the Planck distribution, the Wien displacement,
and the Stefan-Boltzmann laws for black-body radiation (BBR) for a
zero curvature space. Later, Menon and Agrawal \cite{key-12} modified
the $n$ dimensional Stefan-Boltzmann constant found by Landsberg
and de Vos by using the appropriate spin-degeneracy factor of the
photon without affecting normalized Planck spectrum given by Landsberg
and de Vos. Shortly thereafter, Barrow and Hawthorne investigated
the behavior of matter and radiation in thermal equilibrium in an
$n-$dimensional space in the early universe, in particular they calculated
the number of particles $N$, the pressure $p$ and the energy density
$u$ \cite{key-36}. More recently, Gonzalez-Ayala et al. \cite{key-1}
calculated several thermodynamic potentials for BBR such as the Helmholtz
potential $F$, the enthalpy $H$, the Gibbs potential $G$ and the
entropy $S$ by means of the generalized Planck distribution for an
$n-$dimensional Euclidean space. Moreover, they calculated the corresponding
densities per hypervolume unit for these potentials; that is $f$,
$h$, $g$ and $s$ respectively. As it is expected, $g$ and $G$
resulted equal to zero for any spatial dimensionality $n$, such as
it happens for $n=3$. The rest of the potential densities $s$, $f$,
and also the internal energy density $u$ are functions depending
on the absolute temperature $T$ and the dimensionality $n$. When
all these functions were plotted against $T$ and $n$ they seemingly
did not show any interesting feature \cite{key-1}. However, when
a zoom was made in the region of very high temperatures (of the order
of Planck\textquoteright s temperature $T_{P}$) and low dimensionality
(in the interval $n\epsilon(1,10)$), a very surprising behavior was
observed \cite{key-1}. On the other hand, it is well established
that at early times, long before the time of radiation-matter equality,
the universe could have been well described by a spatially flat, radiation-only
model \cite{key-4}. However, in times very close to the Big Bang
within the Planck scale, it is believed that the realm of string theory
\cite{key-5} or loop quantum gravity (LQG) \cite{key-6} is found,
where relativistic quantum gravity effects can be very important.
In the present letter we analyze some thermodynamic properties of
a universe dominated only by radiation within a flat spatial geometry,
beginning at the Planck scale and until times around $t=10^{11}s$,
the time separating the radiation-dominated epoch from the matter-dominated
epoch \cite{key-7-1}. Our analysis is based on an n-dimensional Euclidean
space filled with BBR. As it was shown in \cite{key-1}, this kind
of system exhibits convexities for certain thermodynamic potential
densities only in the region of very high temperatures and low dimensionality.
In particular, the internal energy density $u$, and the entropy density
$s$ show in the first place (as $n$ increases) a local maximum and
then a local minimum while in the Helmholtz free energy density first
a local minimum and then a local maximum. It is very interesting that
the first function that presents a critical point is the Helmholtz
potential density. These extreme points are found within the context
of isothermal processes in a plane of the corresponding potential
density against the dimensionality $n$. Remarkably, the form of the
isotherms is reminiscent of the isotherms found in some first order
phase transitions in gas state equations of the Van der Waals type,
for example (see Fig. 2). If we start at Planck temperature $T_{P}$,
the density $f$ finds its minimum value at $T^{*}=0.93T_{P}$, and
the other two densities ($u$ and $s$) present their extreme points
after this value. If $n$ is considered as an integer, the minimum
of $f$ is located at $n=3$. For any later moment, after the time
corresponding to $T^{*}=0.93T_{P}$, the spatial dimension $n=3$
remains ``frozen''. This extreme point occurs in the unique isotherm
corresponding to a saddle point in the plane $f$ vs $n$. Below this
point a continuous transition from $n=3$ to any $n\neq3$ is forbidden
by the second law of thermodynamics and for all $T>T^{*}=0.93T_{P}$,
any value of $n$ is permitted (allowing for proposals of the type
of string theories \cite{key-5} or even of the type of the so-called
vanishing dimensions models \cite{key-39,key-40,key-41}). This kind
of thermodynamic behavior offers a possible mechanism for determining
the $3+1$ dimensional nature of the space-time.

The question of why is space $3-$dimensional goes back to ancient
Greece \cite{key-37}. In modern times this question was first raised
by Kant in 1746 \cite{key-25}. Later, Ehrenfest in 1917 by means
of the stable orbits argument showed that $n=3$ \cite{key-26}. In
1983, Barrow brought forward a very interesting approach to the dimensionality
problem \cite{key-37}. Since then, many authors have worked on this
problem extended to the case $\left(3+1\right)$; that is, including
time. Such is the case of Brandenberger and Vafa \cite{key-42}, that
in 1989 proposed a natural mechanism for explaining why there are
$3$ large space dimensions in the context of string gas cosmology
\cite{key-42,key-43}. Within the context of the $\left(3+1\right)$
problem Tegmark published an enlightening article summarized through
his Fig. 1 \cite{key-27}. For a deeper discussion on this issue one
can also see the following works \cite{key-38,key-27,key-29,key-30}.

For a black body in an $n$-dimensional space it is known \cite{key-11,key-12,key-36,key-1,key-22,key-23}
that the number of modes per unit frequency interval is,\textcolor{black}{
\begin{equation}
\frac{2\left(n-1\right)\pi^{\frac{n}{2}}V\nu^{n-1}}{\Gamma\left(\frac{n}{2}\right)c^{n}},\label{eq:mod norm vib}
\end{equation}
w}here $V$ is the hypervolume and $n$ the dimensionality of the
space. The multiplication of the above equation by the Bose-Einstein
factor, which is the probability of finding a photon with a frequency
$\nu$ gives the number $dN$ of photons in the frequency interval
$\nu$ to $\nu+d\nu$, 
\[
dN=\frac{2\left(n-1\right)\pi^{\frac{n}{2}}V\nu^{n-1}}{\Gamma\left(\frac{n}{2}\right)c^{n}\left(e^{h\nu/kT}-1\right)}d\nu.
\]

The integration of $h\nu dN/V$ for all the frequencies gives the
energy density (per hypervolume unit) $u\left(T,n\right)$, which
is \cite{key-36,key-1},
\begin{equation}
u\left(T,n\right)=\frac{2\left(n-1\right)\pi^{\frac{n}{2}}\left(kT\right)^{n+1}\zeta\left(n+1\right)\Gamma\left(n+1\right)}{c^{n}h^{n}\Gamma\left(\frac{n}{2}\right)}.\label{eq:energia vol}
\end{equation}

The Helmholtz function\textcolor{black}{{} $F$, t}he entropy $S$,
the pressure $p$ and the Gibbs free energy $G$\textcolor{black}{{}
are given by the following expressions \cite{key-1}
\begin{equation}
F=-\frac{V}{n}u,\; S=\left(\frac{n+1}{n}\right)\frac{Vu}{T},\; p=\left(\frac{1}{n}\right)u,\; G=0,\label{eq:helm-1}
\end{equation}
}respectively. Since the Helmholtz potential density will become very
important in our analysis, the complete form will be given \cite{key-1},
\[
f\left(T,n\right)=-\frac{2\left(n-1\right)\pi^{\frac{n}{2}}\left(kT\right)^{n+1}\zeta\left(n+1\right)\Gamma\left(n+1\right)}{n\; c^{n}h^{n}\Gamma\left(\frac{n}{2}\right)}<0.
\]

The corresponding thermodynamic potential densities are immediately
obtained by dividing each quantity of Eq. (\ref{eq:helm-1}) by the
hypervolume $V$. All these results agree with the well-known 3-dimensional
cases \cite{key-17}. 

In a previous work \cite{key-1} we showed that the thermodynamic
potential densities have regions of minima and maxima when we consider
isothermal processes. In Figure 1 the critical points of $u\left(n,T\right)$,
$f\left(n,T\right)$ and $s\left(n,T\right)$ with respect to the
dimension $n$ ($\partial_{n}u,f,s=0$) are shown. These critical
points can be minima, maxima or saddle points. Notice that for each
function there is a certain critical temperature $T^{*}$ at which
for any $T>T^{*}$ there are no more extreme points of the corresponding
thermodynamic potential density. The region of transition between
such convexities (from maxima to minima or viceversa) occurs at dimensionalities
in the interval from $3$ to $5$ and at very high temperatures (nearly
the Planck temperature). In each case (see Fig. 1) the boundary between
the maxima and the minima is the saddle point corresponding to the
unique temperature $T^{*}$. In Fig. 1 it can be seen that by decreasing
the temperature from $T=T_{P}$, the first function that reaches a
critical point is $f\left(n,T\right)$. The isotherm that passes by
this point is that with $T^{*}\approx0.93T_{P}$. For temperatures
below this value there are maxima and minima of the Helmholtz density
function and it is possible to talk about thermodynamic optimization
criteria. That being said, the first function to optimize was the
Helmholtz free energy density and it happens in a region near the
value $n=3$. This might not be a mere coincidence as will be pointed
out later.

\begin{figure}[H]
\noindent \centering{}\includegraphics{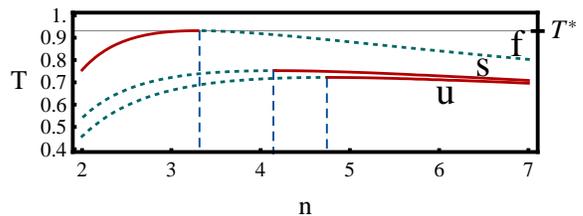}\protect\caption{$\frac{\partial}{\partial n}f,s,u=0$. The regions of maxima and minima
are well defined. The red continuous lines are the minima and the
dotted lines are the maxima of $f$, $s$ and $u$. The blue vertical
dashed lines are located at the unique saddle point of each potential.}
\end{figure}

The separations between the regions of maxima and minima for $u\left(n,T\right)$,
$f\left(n,T\right)$ and $s\left(n,T\right)$ are well defined (see
Fig. 1 and it is also summarized in Table 1 of Ref. \cite{key-1}).

Since the works by Kaluza and Klein \cite{key-7,key-8}, many proposals
about universe models with dimensionality different from three have
been published \cite{key-11,key-12,key-36,key-39,key-40,key-41,key-42,key-43,key-7,key-8,key-32,key-33,key-34,key-35,key-15,key-13}.
Remarkably this has been the case of results stemming from the string,
D-branes and gauge theories \cite{key-42,key-43,key-32,key-33,key-34,key-35,key-15}.
However, nowadays we only have evidence for a universe with three
space and one time dimensions. 

Temperatures as high as $10^{32}K$ are only possible in very early
times in the evolution of the universe. It is known that this period
was dominated by energy in the form of radiation, which means, that
in this period the universe could have been a flat space \cite{key-18}.
Thus, considering the whole primeval universe as a black body radiation
system in an Euclidean space is in principle a reasonable approach.
It is considered \cite{key-20,key-21} that the critical energy density
at the Planck epoch ($t\approx1t_{P}\approx10^{-44}s$, a Planck time)
was around $u\approx1E_{P}/V_{P}\approx10^{9}J/V_{P}$ (an energy
of Planck in a Planck volume) and the temperature around $T\approx1T_{P}\approx10^{32}K$
. The temperature and energy density obtained when the maximum of
the Helmholtz density is located at $n=3$ (in fact $n\approx3.3$,
see Fig 2) are surprisingly close to these values. For example, according
to string theories at some point (at the end of the Planck epoch)
the rest of the dimensions collapsed (or they simply stayed at the
same size) and only the $3-$dimensional space grew bigger \cite{key-5,key-18}.
The remaining question is: why did only $3$ dimensions expand? If
thermodynamic laws were born with the universe, then, thermodynamic
analysis could give us a clue that a maximum or a minimum criteria
would have been fulfilled at the earliest period of our universe.
The conditions of energy and temperature in that epoch were probably
suitable to maximize some kind of thermodynamic function.

In Figure 2 we depict the isotherm curves corresponding to the negative
of the Helmholtz potential density. In fact, from Figure 1, as the
temperature decreases from $T=1T_{P}$ to zero, the first potential
that reaches a critical point (a saddle point) is the Helmholtz energy
density. The first isotherm that reaches a critical point is the one
with $T^{*}=0.93T_{P}$. Between $T=1T_{P}$ and $T^{*}$ there are
no other critical points. After this moment, that is, for temperatures
in the interval from $T<T^{*}$ to $T=0$, there is always first a
maximum and then a minimum of this function. This kind of behavior
resembles the form of the isotherms of a gas-liquid phase transition.
As is well known the critical isotherm (CI) corresponds to a saddle
point which divides the pressure-volume plane in two regions: above
CI, the transitions do not occur and below CI, the transitions are
permitted. However, as we shall see in the present case, in the region
below the critical isotherm $T^{*}$ restrictions imposed by 2nd law
appear and transitions between dimensions are forbidden. With this
in mind, in Figure 2 we can see the isotherms as possible transitions
in the space dimensionality.

\begin{figure}[H]
\selectlanguage{spanish}%
\noindent \centering{}\includegraphics[width=8cm]{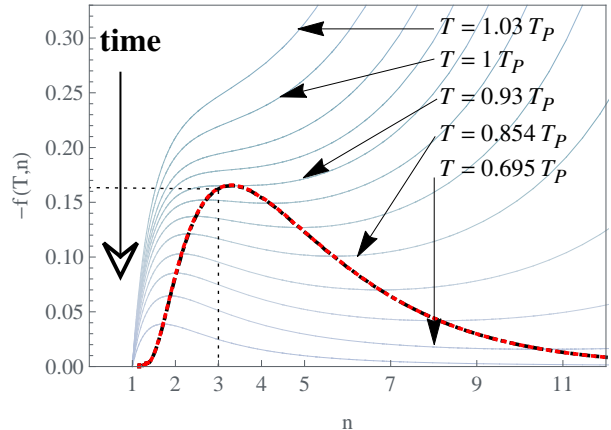}\foreignlanguage{english}{\protect\caption{The continuous line are the isothermal processes, and in dotted line
the critical points. The highest extreme point corresponds to the
saddle point located at the isotherm $T=T^{*}$, above it there are
no more extreme points.}
}\selectlanguage{english}%
\end{figure}

It is very interesting to observe that the critical isotherm $T^{*}$
in Fig. 2 divides the plane $f-n$ in two regions; above $T^{*}$
there are no restrictions over any particular value of $n$. However,
as mentioned before, below $T^{*}$ there will be restrictions on
transitions from $n=3$ to any other dimensionality. 

A well known theorem derived from 2nd law linked with the Helmholtz
potential $F$ is the Helmholtz potential minimum principle \cite{key-24}.
This theorem implies that for isothermal processes
\begin{equation}
\left(\frac{\partial F}{\partial V}\right)_{T}\leq0.\label{eq:condition 2nd law}
\end{equation}

where the equality is fulfilled by reversible processes. Let's consider
a hypercube with volume $V=R^{n}$ as the black body system, where
$R$ is the length of the edge. The above equation can be expressed
as,
\[
\frac{\partial F}{\partial V}=\frac{\partial fV}{\partial V}=f+V\frac{\partial f}{\partial n}\frac{\partial n}{\partial V}\leq0,
\]
where it has been used the consideration that $\frac{\partial f}{\partial R}=0$.
Then,
\begin{equation}
\frac{\partial\left(-f\right)}{\partial n}\geq f\; ln\; R.\label{eq:restriccion 2a ley}
\end{equation}

In an adiabatic expansion (no heat exchange between the universe and
the ``exterior'', a reasonable assumption already considered in
standard cosmology), the condition showed in \cite{key-1} is $pV^{\frac{n+1}{n}}=constant$.
By using Eq. (\ref{eq:energia vol}) and the expression for $p$ in
Eq. (\ref{eq:helm-1}), the adiabatic condition for two different
moments is
\begin{equation}
R=\frac{T_{0}R_{0}}{T}=\frac{T_{P}l_{P}}{T},\label{eq:cond adiabat}
\end{equation}
where $R_{0}$ and $T_{0}$ are some initial conditions that, in our
case, are very near the Planck epoch and $l_{P}\approx1.6\times10^{-35}m$
is the Planck length. This is a result that agrees with typical treatments
in cosmology (see for example \cite{key-20}). Then, Eq. (\ref{eq:restriccion 2a ley})
is now (in Planck units)
\begin{equation}
\frac{\partial\left(-f\right)}{\partial n}\geq-f\; ln\; T.\label{eq:rest 2a ley adiab}
\end{equation}

The difference between the dot-dashed line in Fig. 2 and the restriction
given by Eq. (\ref{eq:rest 2a ley adiab}) (red dotted line in Fig.
2) is very small. According to Eq. (\ref{eq:rest 2a ley adiab}) for
an isothermal process the region below the dotted line is prohibited.
That is, once the integer value $n=3$ is reached at $T^{*}$, for
any $T<T^{*}$ the space dimension of the BBR system remains frozen
at this dimension because, as it can be seen in Fig. (2), any transition
between $n=3$ and $n\neq3$ is not permitted by 2nd law.

As mentioned before, the first objective function reaching an extreme
point was $f\left(n,T\right)$, but is it an adequate quantity to
optimize? Notice that when the internal energy is minimized it is
possible to have an infinite number of configurations, some with more
order than others for the same value of the energy. On the other hand,
by maximizing the entropy it is possible to obtain an endless number
of energetic configurations, thus, by themselves neither the maxima
nor the minima of $s$ and $u$, respectively, determine an advantageous
and unambiguous objective function. This is not the case for $f=u-Ts$,
which gives a kind of trade off between entropy (organization) and
energetic content, and this could be indeed a more meaningful optimization
criterion. When maximized, it offers the better configuration that
gives the less amount of disorder for a given energy. Lets recall
that in stability analysis for closed systems, the stability points
are found in the maximum value of entropy or in minimum energy configurations.
Each one on their own do not establishes an optimum criterion, because
a restriction over $u$ (or $s$) allows an infinite number of configurations
of $s$ (or $u$, respectively). Then, the minimum of the Helmholtz
free energy gives a good commitment between the organization of a
system and its energetic content. In the scenario where the universe
cools down through an adiabatic free expansion, the first potential
whose optimization is reached is the $-f>0$ function. The first maximum
appears at $n=3$. We emphasize the fact that the only isotherm with
a saddle point ($\partial_{n}f=0$ and $\partial_{nn}f=0$) is that
with $T=.93T_{P}$, which could have fixed globally or locally the
dimensionality of the black body system to $n=3$. After this point,
the adiabatic cooling process continues due to the expansion. Nevertheless,
any isothermal dimensionality ``phase'' transition is forbidden
as soon as the temperature diminishes, because for small times $\delta t$
after, the change from $n=3$ to any other value of $n$ is inside
the forbidden zone. It is possible to see in Fig. 1.b that the saddle
point of $f\left(n,T\right)$ is not located exactly in $n=3$. This
could be a consequence of having a simplified model. A possible correction
using the results of quantum gravity can be incorporated by means
of the modified dispersion relation \cite{key-6} stemming from LQG
theories. In ref. \cite{key-1} we found the modified Helmholtz free
energy density $f$ by taking into account the modified dispersion
relation given in \cite{key-6}, obtaining the following result \cite{key-1},
\[
f\left(n,T\right)=-\frac{2\pi^{\frac{n}{2}}\left(n-1\right)T^{n+1}}{\Gamma\left(\frac{n}{2}\right)}\left\{ \frac{\Gamma\left(n+1\right)\zeta\left(n+1\right)}{n}+\right.
\]
\[
\frac{\left(\frac{3}{2}+\frac{n-1}{2}\right)\alpha T^{2}\Gamma\left(n+3\right)\zeta\left(n+3\right)}{n+2}+\left[\alpha^{2}\left(-\frac{5}{8}+\frac{3\left(n-1\right)}{4}\right)+\right.
\]
\begin{equation}
\left.\left.\alpha'\left(\frac{5}{2}+\frac{n-1}{2}\right)\right]\frac{T^{4}\Gamma\left(n+5\right)\zeta\left(n+5\right)}{n+4}\right\} ,\label{eq:u modificada grav cu=0000E1ntica-1}
\end{equation}
where $\alpha$ and $\alpha'$ take different values depending on
the details of the quantum gravity candidates. By manipulating the
value of the parameters $\alpha$ and $\alpha'$ in Eq. (\ref{eq:u modificada grav cu=0000E1ntica-1}),
which depend on the LQG model, it is possible to modify the value
at which the saddle point occurs. In Fig. 3, it is depicted the analysis
of maxima and minima for the Helmholtz potential density incorporating
the LQG corrections. A similar analysis of critical points reveal
that the corresponding saddle point $T^{*}$ decreases in one order
of magnitude.

\begin{figure}[H]
\noindent \centering{}\includegraphics{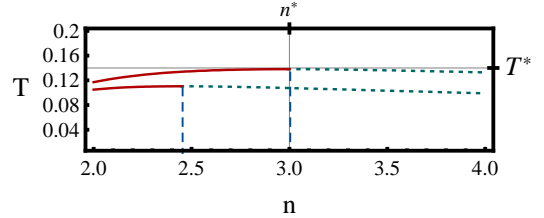}\protect\caption{Maxima (dotted lines) and minima (continuous lines) of $f$ incorporating
LQG corrections. In the case $\alpha=0.1$ and $\alpha'=2.28$ (these
values reproduce the curves of the spectral energy density showed
in \cite{key-6}) the saddle point is located at $n\approx2.44$.
In the case $\alpha=0.15$ and $\alpha'=.001$ the saddle point is
at $n\approx3$ (see vertical shaded lines).}
\end{figure}

In this way it has been proposed a model based on simple suppositions
to study scenarios that put the space dimensionality $n\approx3$
as a convenient candidate to optimize a thermodynamic quantity: the
Helmholtz potential density. After picking this value, later transitions
were prohibited by second law restrictions and the dimensionality
remained ``frozen'' at $n=3$. In regard the $1$ in $\left(3+1\right)-d$,
is generally accepted that the arrow of time was imposed by the principle
of entropy increment for closed systems. Thus, possibly the laws of
thermodynamics selected the dimensions of the universe.
\begin{acknowledgement*}
We want to thank partial support from COFAA-SIP-EDI-IPN and SNI-CONACYT,
M\textsl{\'E}XICO.
\end{acknowledgement*}

\end{multicols}{}
\end{document}